\documentclass[aps,preprint,a4paper,prc,showpacs,showkeys,
nofootinbib]{revtex4-1}
\usepackage{graphicx}
\usepackage{bm} 
\usepackage{amssymb,amsmath,amsfonts}

\begin{document}

\title{Centrifugal effects in $N\Delta$ states}
\author{J.A. Niskanen}

\affiliation{Helsinki Institute of Physics, PO Box 64, 
FIN-00014 University of Helsinki, Finland }
\email{jouni.niskanen@helsinki.fi}
\date{\today}

\begin{abstract}
Recently it has been pointed out that in the 
two-baryon $N\Delta$ or $\Delta\Delta$ system the width of
the state is greatly diminished due to the relative kinetic 
energy of the two baryons, since the internal energy
of the particles, available for pionic decay, is smaller.
For nonzero orbital angular momenta this effect becomes
state dependent. Also the real part of the $N\Delta$ energy
can get contribution from the centrifugal barrier leading to
rotational series of diffuse states. Obviously, these states
may have an interpretation as isospin one dibaryons. Direct
and explicit calculations for this are presented with some 
details of the coupled state wave functions displayed. With 
finite expectation values of these repulsive effects it may be
possible to define state dependent effective thresholds for
$N\Delta$ states and these, in turn, can show resonant like
behavior.
\end{abstract}

\pacs{ 25.80.-e, 21.85.+d, 13.75.-n, 24.10.Ht}
\keywords{dibaryons, resonances}

\maketitle

\section{Introduction}
The role of the $\Delta(1232)$ resonance has a long and 
established history in the pion-nucleon interaction and also
in the $NN$ interaction and plays a outstanding part
in their inelasticities \cite{green,arenhovel}. Pionic
inelasticities would be even idiomatic for probing
intermediate $N\Delta$ components arising from incident
$NN$ states above the $\pi NN$ threshold.  Isovector 
meson, pion, exchange (and also $\rho$ exchange) can excite an
$N\Delta$ state, which in turn decays. The process is inseparable
from elastic $NN$ scattering and their strong interaction.

As the simplest such reaction, $pp \rightarrow d\pi^+$ was
intensely studied in the late 70's and 80's at ``meson 
factories'' SIN, LAMPF and TRIUMF. For a modern review see
e.g. Ref.~\cite{hanhart}. Experiments
indicate univocally the $\Delta$ peaking in the total
cross section at about 580 MeV laboratory energy
\cite{wedpla,wednpa}, which, along with  the differential 
cross section and transverse analyzing power, were also 
reasonably predicted by a coupled channel $N\Delta$ 
calculation without virtually any free
parameters~\cite{ppdpi}\footnote{A form factor to account
for the nucleon and $\Delta$ size could be considered as a free 
parameter, though it was taken from OBE potentials.}. Further,
spin correlations and deuteron polarization were given 
predictions~\cite{double,dpolar}, which were promptly 
confirmed~\cite{aprile1,aprile2}.
The success of the calculations was mainly due to two
$N\Delta$ admixtures generated from the initial 
nucleons:
$^1D_2(NN) \rightarrow {^5S_2}(N\Delta)$ and
$^3F_3(NN) \rightarrow {^5P_3}(N\Delta)$ sometimes dubbed as
``dibaryons''\footnote{A third essential ingredient
was the axial charge and $s$-wave pion rescattering in the
$^3P_1(NN)$ and associated $N\Delta$ waves.}. 
The latter was
probably the first appearance of such a high partial wave 
appearing as important \cite{ppdpi}. It may be noted
that the coupled channel $N\Delta$ agreement with experiment 
was actually better than a six-parameter dibaryon 
fit~\cite{kamo}. 

Later an improvement~\cite{improv} was done, which decreased 
the widths of the $N\Delta$ states, so far normally 
considered as the free $\Delta$ width with just appropriate 
kinematic adjustment~\cite{sainio,leidemann}. 
Namely, there is relative 
kinetic energy in the $N\Delta$ intermediate
state, whose role should be considered in more detail.
Obviously this kinetic portion is not usable for the 
(internal) pionic decay of the $\Delta$'s. 
The width in turn is used as a uniform imaginary potential
in the coupled $N\Delta$ system. 
Further, because the $N\Delta$ wave function becomes now 
necessarily spatially constrained (must vanish asymptotically) 
both below and also above the $N\Delta$ threshold, 
this kinetic energy is not arbitrary. Its average is finite
and can be calculated from the Fourier transform of the
wave function. The resulting kinematic suppression of the
width was taken into account in the calculations of 
Ref.~\cite{improv} for $pp \rightarrow d\pi^+$, but explicit
width results were only published recently~\cite{delta2}. 
Also rather strong sensitivity of the width was seen on 
the relative orbital angular momenta of the baryons.
In addition to the improved agreement in the old reaction
$pp \rightarrow d\pi^+$, 
the reduction of the width has great renewed relevance for 
interpretations of the $I(J^P) = 0(3^+)$ resonance recently
discovered  at the WASA@COSY detector of Forschungszentrum 
J\"ulich and labeled as $d'(2380)$ 
dibaryon~\cite{adlar106,adlar721}. Its nominal
mass 2380~MeV is 80~MeV below the $\Delta\Delta$ threshold and
its width only 70~MeV, less than that of a free $\Delta$. 
A calculation~\cite{delta2}, similar to that shortly described
above, gives just such a decreased value for the 
$^7S_3(\Delta\Delta)$ state coupled to $^3D_3(NN)$.
A similar drastic suppression of the double $\Delta$ system 
width is also found by Gal and Garcilazo~\cite{gal1,gal2}
from coupled pion-baryon Faddeev calculations.

Further, in these reactions and $NN$ scattering in different partial 
waves also the $N\Delta$ centrifugal barrier directly
diminishes the wave functions. Although this suppression is 
expected to be naturally sensitive to $L_{N\Delta}$, also the 
orbital angular momentum of the initial nucleons may even
favor transitions into $N\Delta$ in some sense.
Namely, within the interaction range the reduction of the 
centrifugal barrier can compensate the $N\Delta$ mass 
difference in the excitation if $L_{N\Delta} < L_{NN}$,
as seen in Ref.~\cite{dibarmass} as an explanation for $T=1$
enhancements ($T=1$ ``dibaryons''). From the above
considerations it is clear that just
a single number cannot account for the effective two-baryon
pole position in different partial waves.

Because the internal kinematics of the $N\Delta$ system has
been seen to have  a significant effect on its width, 
i.e. the imaginary part of the interaction~\cite{delta2}, 
it is also of interest to see the effect
on the real part of the energetics.
Due to the confinement of the $N\Delta$ wave function, 
similar to bound systems in quantum mechanics, one might expect 
some kind of imitated quantization of the energy to appear - 
as noted above about the finite distribution of kinetic energy. 
In this kind of situation the angular momentum is directly
related to the energy and
this aspect of the kinetic energy is our central point now.
Thus, if the expectation value of the centrifugal barrier
$\hbar^2 / 2mr^2$ is well defined
and reasonably constant over various $N\Delta$ configurations
and $NN$ energies one might expect also a reasonably well
defined rotational series $\propto L_{N\Delta}(L_{N\Delta}+1)$
to appear as effective channel thresholds. This is the direction
to proceed now.

In fact, a very phenomenological calculation~\cite{dibarmass}
gave some hints for this possibility. The work compared a
phenomenological phase-fitted potential~\cite{reid}
and the same potential
supplemented by an $N\Delta$ channel. To remove the double 
counting of attraction due to the extra $N\Delta$ component, 
a repulsion $[V_2(r)]^2 / \Delta E$ was added
to mimic in a closure approximation the second order $\Delta$
effect. Here $V_2(r)$ is the $NN \leftrightarrow N\Delta$
transition potential and $\Delta E$ an average energy 
denominator adjusted for phase equivalence of the coupled and
uncoupled calculations. It was seen that $\Delta E$ followed 
quite well such a rotational series with 
$\hbar^2 / 2mr^2 \approx 40$~MeV
indicating about 1 fm as the effective channel radius. A
remarkable thing was that this pattern 
actually corresponded well the series of isospin one dibaryons 
reported in Refs. \cite{yokosawa,yokopolar}. An additional 
criterion for a ``dibaryon'' to appear in the $N\Delta$ 
series was that $L_{N\Delta} < L_{NN}$, meaning that in the 
$NN \rightarrow N\Delta$
transition the orbital angular momentum decreases. For short
range energetics this is a natural expectation, since then
the $\Delta - N$ mass barrier is partly compensated by massive
reduction of the centrifugal barrier favoring the tunneling into
the $N\Delta$ channel.

The aim of this paper is to study in a simple transparent 
way some phenomenological aspects of how the relative kinetic 
energy between the two intermediate baryons influences the 
overall dynamics of the two-baryon system and, in particular, 
to calculate explicitly the
expectation value of the centrifugal barrier for realistic
coupled channel $N\Delta$ wave functions.
First, in Sec.~\ref{formalism} a brief review is given
about the formalism before proceeding to calculations and
results in~\ref{results}.

\section{Formalism}   \label{formalism}
Pionic excitation of the $N\Delta$ (and $\Delta\Delta$) components
into $NN$ configurations was already suggested in an early
paper by Sugawara and von Hippel \cite{suga}.
The coupled channel formalism is generally described in 
the reviews \cite{green,arenhovel} and relevant details of the 
interactions are provided in Ref.~\cite{ppdpi}. Also essential
updates and improvements are given in Ref.~\cite{improv}
mainly intended for the reaction $pp \rightarrow d\pi^+$, but
extending relevantly in $V_2$ and in the width
for more general $N\Delta$ dynamics. In particular,
the peak of this reaction may be the best constraint on its
dominant transition potential $V_2$. The structure
can be presented routinely by the coupled radial 
Schr\"odinger equation
\begin{eqnarray} 
&   \displaystyle \left[ \frac{\hbar^2}{2m} 
\left( \frac{d^2}{dr^2} - \frac{L(L+1)}{r^2}\right)
 - V_1(r) + E \right] u(r) = V_2 w(r)  \label{nneq}  \\
&   \displaystyle \left[ \frac{\hbar^2}{2m'} 
\left( \frac{d^2}{dr^2} - \frac{L'(L'+1)}{r^2}\right)
 - V_3(r) + (E - \Delta) \right] w(r)= V_2 u(r) \label{ndeq}
\end{eqnarray}
for the $NN$ and $N\Delta$ wave functions $u(r)$ and $w(r)$
respectively. Here the generic channel potentials are denoted
by $V_i(r)$, and $L$ ($L'$) and $m$ ($m'$) present the
angular momentum and reduced mass of the $NN$ ($N\Delta$)
system. The $\Delta - N$ mass difference $M_\Delta - M_N$
is abbreviated as $\Delta$, with $M_\Delta$ taken as the position
of the pole 1310 MeV rather than the nominal mass~\cite{pdg}. 
Above pion production threshold it is complemented by the state 
dependent width~\cite{delta2} into the form $\Delta - i\Gamma /2$.
Thus, effectively the width acts as a constant imaginary part 
in the potential making the $N\Delta$ channel wave function $w(r)$
asymptotically
vanishing. The asymptotically free radial $NN$ wave function 
with momentum $k$ can be presented and normalized with
the descriptive form $u_{NN}(r)\sim kr\, \exp (i\delta_L) \, 
j_L(kr+\delta_L)$, where the phase shift may now be complex. For 
real interactions and below open channels it is common to the
total wave function (with all channels). As a numerical comment,
one should remember the rather slow asymptotic convergence 
of the regular and irregular Bessel function 
combination~\cite{taylor} 
$u_{NN}(r)\sim kr\, \exp (i\delta_L) \, 
[ \cos\delta_L \,j_L(kr) - \sin\delta_L\, n_L(kr)]$
to this form for $L \neq 0$.
The inclusion of more possible channels in the system
(\ref{nneq}) - (\ref{ndeq}) is obvious.

Also it is interesting to note the emergence of inelasticity above
the $\Delta - N$ threshold even without any imaginary potential.
With the opening of the $N\Delta$ channel unitarity in the $NN$
sector is lost. Similar to the case with the tensor force, then 
the single parameter, the $NN$ phase shift, is not sufficient
any more to describe the asymptotics but a transition amplitude
(analogous to $\epsilon_J$) and scattering of the $N$ and the
$\Delta$ would be necessary. Below this threshold it is still
possible
to desribe the scattering wave functions as real (by switching
off the common overall phase $\exp{(i\delta_L)}$). In this case
the still closed $N\Delta$ channel wave function $w(r)$ behaves
asymptotically like $\exp{(-\kappa r)}$ with 
$\kappa = \sqrt{2m'(\Delta - E)}/\hbar$. Apparently at $N\Delta$
threshold the extension of the wave function becomes very large 
producing a cusp in the $NN$ phase shift. Also the overlap
integral of the amplitude in e.g. $pp \rightarrow d\pi^+$
would maximize grossly overestimating the cross section.
Above threshold
the phases (arguments) of both wave functions $u(r)$ and
$w(r)$ depend on the channels and even on the radius $r$. 
In this case the
$N\Delta$ wave function $w(r)$ behaves like an outgoing spherical
wave $\propto \exp{(ik'r)}$ or more accurately
$k'r\,h^{+}_{L'}(k'r)$ with $k' = \sqrt{2m'(E-\Delta)} /\hbar$. 
The oscillatory behavior decreases overlap integrals from the
cusp peak values.
Probably in this case it would be possible to define scattering
eigenfunctions and corresponding phase shifts like those of
Blatt and Biedenharn \cite{blatt}.
However, in the presence of the constant imaginary potential 
$-i\Gamma /2$ this parametrization is not useful. In this case
complexity appears already below the nominal channel threshold
adding to the damping and bringing in some oscillatory behaviour.
Just at threshold the asymptotic suppression 
from the width should behave like $\sim \exp (-\gamma r)$ and 
include also corresponding oscillation with wave number
$\gamma = \sqrt{m'\Gamma/(2\hbar^2)}$ (interfering with the
effect of $V_2(r) u(r)$). The width moderates and rounds the 
cusp peak down. Above threshold this damping
effect is sustained along with oscillations
having the natural wave number $k'$ given above.

Among the interactions the most important is the 
transition potential $V_2(r)$,  
based on one pion exchange supplemented with
$\rho$ exchange. Heavy meson exchanges to describe short-range
interactions may not be favoured
in present day effective field theories, but the main point is
that this potential is thoroughly tested in the reaction
$pp \rightarrow d\pi^+$~\cite{improv}
and is now used only to imply the form of
the associated $N\Delta$ component in detail. The diagonal
$V_1(r)$ in the $NN$ sector is taken as the old 
phenomenological Reid potential~\cite{reid}
modified to give the correct phase shifts in the presence of
the $N\Delta$ excitation~\cite{ppdpi,improv,csb}. The $N\Delta$
potential $V_3(r)$ is not in our primary interest presently and 
is neglected. In $NN$ scattering it would only appear hindered
behind iterated $V_2$. The width
is calculated along the lines of Ref.~\cite{delta2}, exhibited
there for the most important and interesting states. In 
addition to the fact of giving rise to inelasticities 
and making also the $NN$ wave function $u(r)$ complex, it causes 
also rather strong repulsion~\cite{etacpl,antip}
decreasing the attractive 
effect due to $N\Delta$ excitation. And it is state dependent
as expressed in the Introduction.

To see the effect of the centrifugal barrier and the possible
appearance of the rotational series one needs the 
straightforward expectation value
\begin{equation} \displaystyle
\langle\, \frac{1}{r^2} \, \rangle = 
\frac{\int_0^\infty |w(r)|^2\,
/r^2\, dr}
{\int_0^\infty |w(r)|^2\, dr} .
\end{equation}
Apparently this is simpler than the calculation 
\cite{improv,delta2} 
\begin{equation}
\Gamma_3 = \frac{2}{\pi}\,
\frac{\int_0^{p_{\rm max}} |\Psi_{N\Delta}(p)|^2\,
\Gamma(q)\, p^2\, dp}
{\int_0^\infty |\Psi_{N\Delta}(r)|^2\, r^2\, dr} 
\label{gamma3}
\end{equation}
for the width
requiring  the Fourier transform of the wave function and
relevant restraints for the kinematically allowed momenta
$p$ and $q(p)$.
The series $\langle \hbar^2 / (2m'r^2) \rangle L'(L'+1)$
should be built on top of the mass difference $\Delta$.
In fact, one can also calculate the still missing   
kinetic energy $\langle \, - \hbar^2/2m' \;
\partial^2/\partial r^2 \, \rangle$, associated with the
radial degree of freedom,
to be added to the nominal threshold $\Delta$. Most naturally 
this can be calculated as the expectation value of $p^2/2m'$ in 
the momentum representation as in Eq. (\ref{gamma3}).

\section{Results}   \label{results}
\subsection{$N\Delta$ component magnitudes}   \label{sizes}
In this section we study first details of the $N\Delta$ components
connected to isospin one $NN$ wave functions for various angular momentum
configurations and then calculate consequent energy expectation values 
relevant as effective channel thresholds.

\begin{figure}[tb]
\includegraphics[width=\columnwidth]{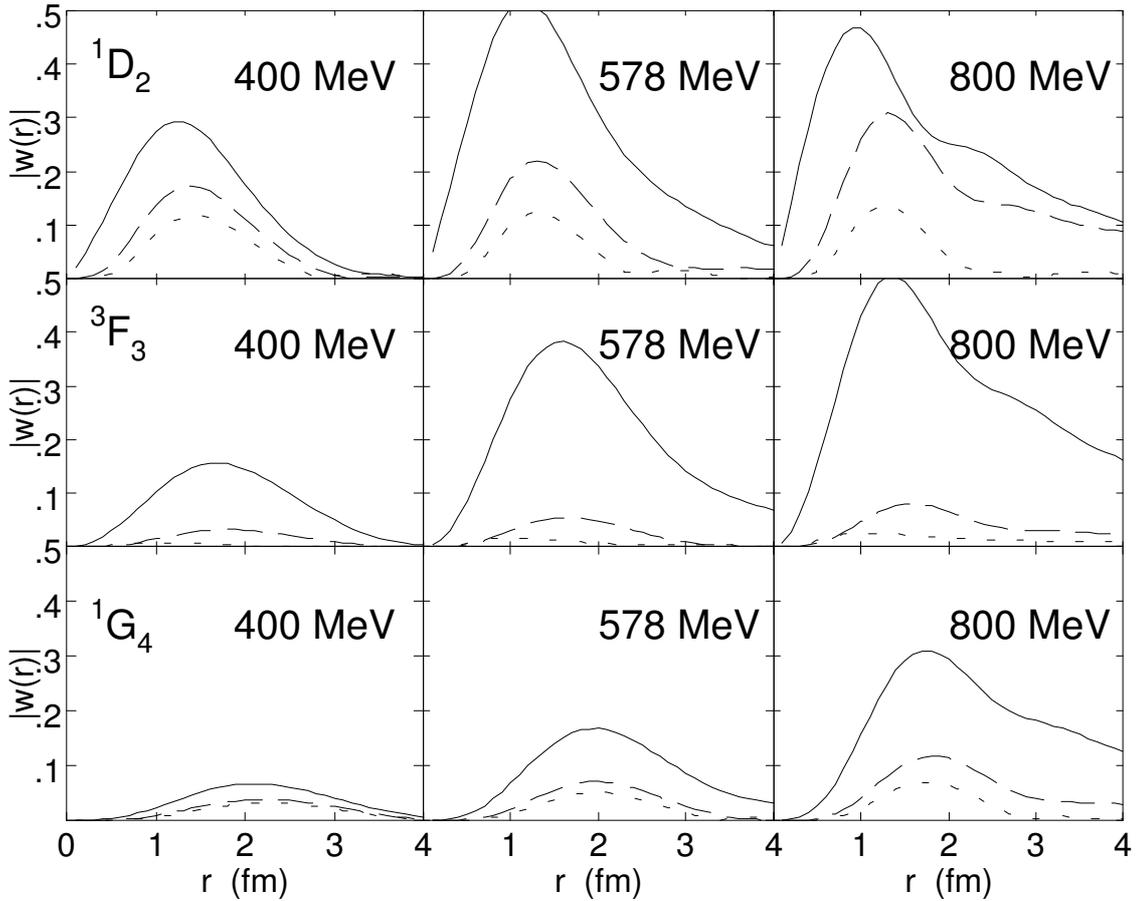}
\caption{Absolute values of the $N\Delta$ wave functions
for the best ``dibaryon'' $NN$ initial states at three laboratory
energies. The curves are explained in the text.} \label{absdiba}
\end{figure}

Figure \ref{absdiba} presents the radial dependence of the
absolute values of the $N\Delta$ wave function components
as described in the previous section. The uppermost functions
present the $^5S_2(N\Delta)$ (solid curve), $^5D_2(N\Delta)$
(dashed) and $^5G_2(N\Delta)$ (dotted) states associated with the 
incident $^1D_2(NN)$ partial wave at three laboratory energies.
This is the lowest lying ``dibaryon'' from $NN$ phase shifts
(apart from the deuteron and the sharp low-energy maximum 
in the quasibound  $NN$ $^1S_0$ wave).
The position of the peak is practically independent of 
energy, for high energies at slightly smaller distance. 
Even up to 700 MeV the dominant $S$-wave maximum remains at 
about 1.1 -- 1.2~fm, and the $D$- and $G$-wave variations from this 
distance are also negligible.\footnote{The odd looking energy 
578~MeV is chosen here, because about at this energy the 
maximum of the experimental total 
cross section of the reaction $pp \rightarrow d\pi^+$ is reached,
which can be used for fixing the $NN \leftrightarrow N\Delta$ 
transition potential strength \cite{wednpa,improv}.} 
Such weak energy dependence was
also observed in Ref.~\cite{dibarmass} in the energy denominator
$\Delta E$ of the second order effective repulsion
$V_2^2/\Delta E$ to cancel the attraction due to the $N\Delta$
coupling and keep phase equivalence of pure single channel $NN$ 
scattering and the coupled model. The repulsive influence of the
centrifugal barrier is clear in the ordering of sizes, even more
notably considering that the relative strengths of the (radially
identical tensor type) transition potentials would be in opposite 
ordering $20:(-23):31$ for  $^5S_2(N\Delta)$, $^5D_2(N\Delta)$ 
and $^5G_2(N\Delta)$, respectively~\cite{ppdpi}. However, in spite of 
this, the position of the maximum is remarkably independent of $L'$.
The shoulders in $r$-dependence are due to either the real or imaginary
part passing through zero. It might be reminded that the present 578~MeV
result is related to the one in Fig. 1 of Ref. \cite{delta2}, where
just the real part of the wave function was presented.

The second clear ``dibaryon'' candidate in $NN$ 
scattering~\cite{yokosawa} and pion production~\cite{ppdpi} appears in 
the $^3F_3(NN)$ initial state
shown in the second row of Fig.~\ref{absdiba}. Again the 
lowest $L' = L-2$ coupled channel angular momentum $^5P_3$ 
(solid curve) is clearly  favoured and the position of its 
maximum is also fairly constant $\approx$1.6~fm
over a wide range of energies. The dashed and dotted curves correspond 
to the much smaller $^5F_3(N\Delta)$ and $^3F_3(N\Delta)$ 
states, respectively. Remarkably, also in this case the effect of 
the $N\Delta$ angular momentum $L'$ has only minuscule influence 
on the position of the maximum, whereas in comparison to the 
above $^1D_2(NN)$ case the {\it initial} $NN$ wave 
being different with $L=3$ has a larger effect. Again at highest 
energies the position of the maximum tends towards smaller distances, 
a behavior to be discussed later. It is also worth noting
that in the case of $^3F_3(N\Delta)$ the tensor and spin-spin like 
parts of the transition potential act mutually destructively making
the $^3F_3(N\Delta)$ wave negligible. The high $^5H_3(N\Delta)$ component
is not included.

A third possible candidate satisfying the ``dibaryon'' 
conditions discussed in the previous sections has already quite a high 
angular momentum $^1G_4(NN)$. The preferred $N\Delta$ angular momentum 
state is now $^5D_4(N\Delta)$, about as distinct as the previous 
$^5D_2(N\Delta)$ as seen in the lowest $L$ series of Fig.~\ref{absdiba} 
(solid curve). The  $^5G_4(N\Delta)$ and  $^5I_3(N\Delta)$ waves 
(dashed and dotted curves) are smaller but not negligible. 
Again the position of the maximum depends on the energy and $L'$ only 
rather weakly, whereas the initial $L=4$ of the initial nucleons has 
pushed the maxima to about 2~fm. The $^5D_4(N\Delta)$ maximum levels 
to $\approx$0.37 around 1100 and 1200~MeV. So the possible peaking 
of the energy distribution would be rather wide in this region.

A common systematic feature in the different states is that 
with increasing energy, above 600 MeV, 
there is a common (albeit weak) tendency  
for the maximum to creep towards smaller distances, since then
also the incident particles get closer to each other through their
centrifugal barrier. Another is that one may well speculate of 
some effective threshold, appearing as a cusp peak which
in the case of $^1D_2(NN)$ is passed around 600~MeV (in $^5S_2(N\Delta)$
and may be reached for $^3F_3(NN)$ at about 800~MeV (in $^5P_3(N\Delta)$), 
whereas for $^1G_4(NN)$ it is still ahead. This expectation is confirmed 
by actual calculations as noted above. It is also worth noting that the
formally calculated widths \cite{delta2} are about 100 MeV at the relevant
energies of maximal wave functions in full agreement with 
Yokosawa~\cite{yokosawa}.

\begin{figure}[tb]
\includegraphics[width=\columnwidth]{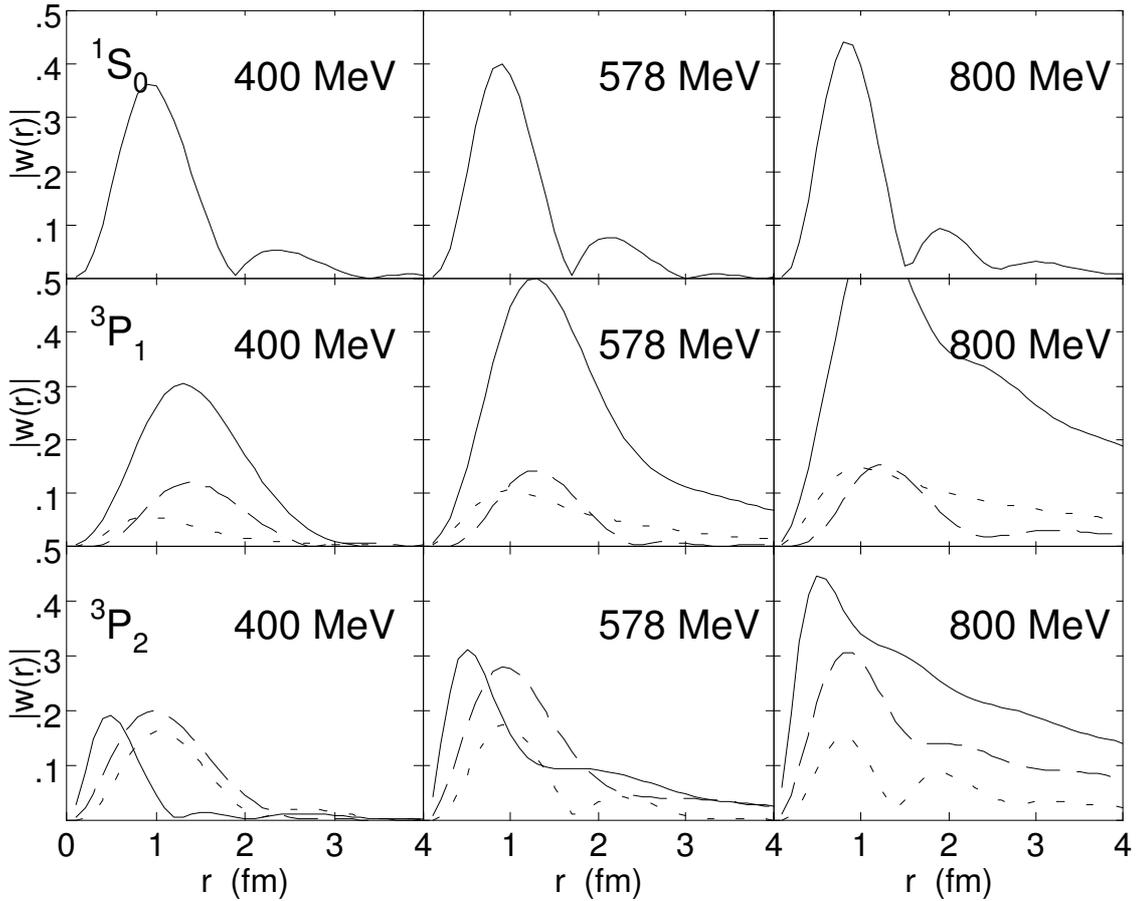}
\caption{Absolute values of the $N\Delta$ wave functions
for the ``bad'' dibaryon $NN$ initial states.  The curves are
explained in the text.} \label{nondibar}
\end{figure}

Now it may be of interest to have a look at states which do {\it not}
satisfy the optimal conditions for ``dibaryons'' discussed earlier.
The first such study may well come from the lowest $L$ state, 
$^1S_0(NN)+ {^5D}_0(N\Delta)$, shown on the first row of 
Fig.~\ref{nondibar}. The magnitude of the wave function is quite
considerable, though it remains practically independent of energy. 
So it is not likely to produce such energy dependent behavior as
resonances.
In fact, the inclusion of the $^5D_0(N\Delta)$ component
into the wave function introduces very strong attraction -- even
tens of degrees into the phase shift $\delta_0$.
At the lowest energy 400~MeV the maximum is just at 1~fm and 
with increasing energy creeps to a slightly closer radius,
to about 0.8~fm at 800~MeV. With its magnitude it looks quite strange 
that this state contributes very little to the reaction
$pp \rightarrow d\pi^+$. A partial reason is that the $NN$ and $N\Delta$
contributions tend to cancel rather completely in the $^1S_0$
amplitude to this reaction. Also, in overlap integrals 
this $N\Delta$ component mainly requires the smaller
$D$-wave component of the deuteron.

The $^3P_1(NN)$ wave has three $N\Delta$ mixing states $^5P_1(N\Delta)$,
$^5F_1(N\Delta)$ and $^3P_1(N\Delta)$ shown next in Fig.~\ref{nondibar}
by solid, dashed and dotted curves, respectively. In this case there is a strong energy dependence in the $P$-wave $N\Delta$ components with 
$^5P_1(N\Delta)$ reaching 0.6 at 800~MeV. The maximum 
(in $r$ variable) levels to this value
at 900~MeV and turns then slowly down, so slowly that the $r$-maximum remains
within 2\% from this value from 800 to 1000 MeV. One might imagine 
a resonance at this energy, but the background $^3P_1$ phase shift is going 
down so fast with increasing energy that perhaps it is not possible to see
any ``dibaryon'' in this wave, in particular since the energy
dependence of the very wide peak may be too slow. 
In the reaction $pp \rightarrow d\pi^+$ this wave gives rise to 
$s$-wave pions and pion-nucleon $s$-wave rescattering 
interferes destructively so that basically the corresponding amplitude 
is quite flat and very small rather soon above threshold, say above 
400~MeV~\cite{ppdpi}. The dotted curve, 
$^3P_1(N\Delta)$, deviates slightly from the behavior of others, 
since it has also a moderate contribution
from the spin-spin type transition potential~\cite{ppdpi}, whereas the 
others arise purely from the tensor type. This coupling stresses
typically shorter ranges.

The $N\Delta$ components in the tensor coupled $^3P_2-{^3F}_2$ $NN$
initial states are somewhat smaller. The lowest row of Fig.~\ref{nondibar}
shows those arising from the state with the initial asymptotic wave 
$^3P_2(NN)$ boundary condition, although  
by the earlier arguments initial $^3F_2(NN)$ might be presumed 
to be dominant. The solid curve shows the $^3P_2(N\Delta)$ component and 
the dashed one  $^5P_2(N\Delta)$. The former now arises dominantly from the
spin-spin coupling, which clearly tends to emphasize smaller ranges than 
the tensor. In spite of its formally two times stronger transition 
potential~\cite{ppdpi} the $^5F_2(N\Delta)$ (dotted) is clearly smaller 
than the $P$-waves and 
the $^3F_2(N\Delta)$ (not shown) again roughly one half of this.
The minima reflect the nodes of the initial partial wave. Lacking these 
nodes at small distances, the $N\Delta$ admixtures in $^3F_2(NN)$ are 
smoother but also smaller than in $^3P_2(NN)$, about half of these 
in height with their maxima $\approx 0.2$ situated at 1.5~fm again 
showing the effect of the initial $L$ to the $N\Delta$ wave function.

\subsection{Phase considerations}     \label{phases}
In some approaches to $N\Delta$ effects (e.g.~\cite{brack} and 
\cite{plato} for pion production) the mixing wave functions are
considered more or less implicitly in a factorization approximation
$w(r) \approx V_2(r) u(r) / (E-\Delta +i\Gamma /2)$. 
Therefore, it may also be of interest to compare this approach
by plotting the phase of the coupled wave function relative to
the $NN$ wave function. Namely, since differential observables, 
in particular polarization phenomena, are sensitive to the phases
of the amplitudes, also this phase may matter in reactions where
$N\Delta$ components are active participants. Further, it should be 
noted that the $L'$ dependence of the width, generated by the wave
function structure \cite{improv,delta2}, is not often included.

This kind of treatment may give the total cross section peaking
by construction relatively trivially \cite{brack}. However, as 
shown e.g. in Ref. \cite{polarconf},  due to the different 
centrifugal barriers even the relative sizes of the $N\Delta$
components $^5S_2 + {^5D_2} + {^5G_2}$ coupled to $^1D_2(NN)$ would
come wrong as pointed out also in Subsec. \ref{sizes}. 
This kind of effects and
and also phase interferences lead to incorrect differential
cross sections \cite{chai,maxwell1} and spin observables
\cite{maxwell2} in $pp \leftrightarrow d\pi^+$. Differential
observables are not tested at all in e.g. the ``hidden dibaryon''
approach to $pp \rightarrow d\pi^+$ of Ref. \cite{plato}. 
It is doubtful that this test would be passed.

\begin{figure}[tb]
\includegraphics[width=\columnwidth]{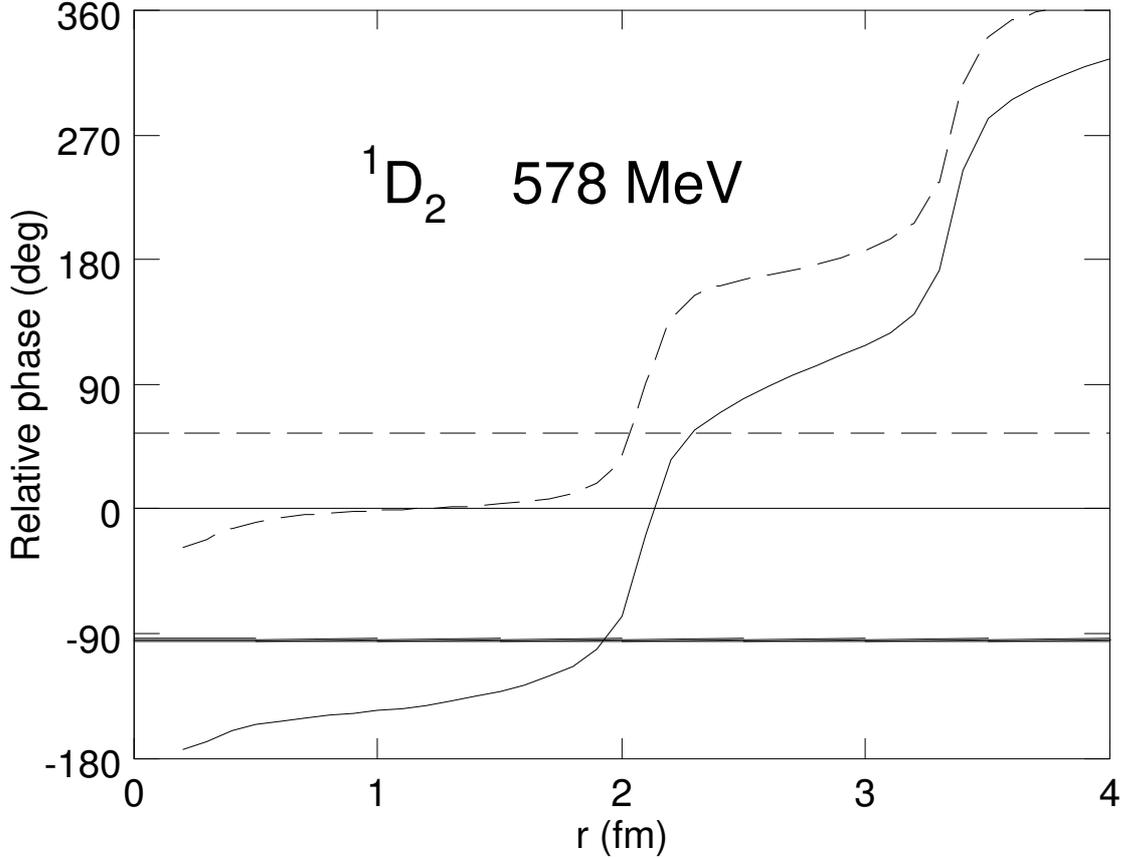}
\caption{Relative phase between $N\Delta$ and $NN$ wave functions
for the best dibaryon $NN$ initial state $^1D_2$ at 578 MeV 
laboratory energy. The solid and dashed
curves present $^5S_2(N\Delta)$ and $^5D_2(N\Delta)$, respectively.
The horizontal lines show the phase of the $N\Delta$ propagator,
independent of $r$, as explained in the text.} \label{relativephase}
\end{figure}

Clearly, in this approximation
division of $w(r)$ by $u(r)$ should rid the wave function from 
the overall nucleonic phase shift $e^{i\delta}$ so that
the resulting function should only have a phase from the 
propagator, independent of $r$, and from the sign of $V_2(r)$.
Fig.~\ref{relativephase} for $^5S_2(N\Delta)$
and $^5D_2(N\Delta)$ admixtures in the $^1D_2(NN)$ state at 
578~MeV shows this simple assumption to be problematic. 
This energy is chosen 
very close to the threshold cusp to maximize the presence of the 
resonant cusp effect. Obviously zero and 180 degree phases
(modulo 360 degrees) would correspond to the same or opposite signs 
between the wave functions, which, in turn, is connected to the sign
of the transition potential $V_2(r)$. Below and close to pionic
threshold, naturally, this remaining relative 
phase alternates between $\pm 180^\circ$
and 0. In the presence of strong inelasticity this behavior is more 
blurred and Fig.~\ref{relativephase} should be understood 
analogously in terms of the complex wave function ``changing signs''. 
So in Fig.~\ref{relativephase} the solid curve below 2~fm corresponds
to positive $V_2(r)$ and positive {\it vs.} negative $NN$ and $N\Delta$
wave functions, and oppositely the dashed curve to negative $V_2(r)$
and positive $NN$ and $N\Delta$ wave functions, until, at 2~fm,
the $NN$ wave function changes sign (or more literally 
$e^{-i\delta} u(r)$ changes sign; with stronger inelasticity even this
would have significant imaginary component in addition to the one
generated in $N\Delta$).
Of course, in actual amplitudes $r$-dependence is integrated
over. 

The phase from the propagator $(E-\Delta -i\Gamma /2)^{-1}$ can be
easily compared. In the comparison one should also be aware that, 
as stated previously, the width is state dependent~\cite{delta2}. 
The straight horizontal lines show the phases arising from the 
propagator of the factorization with three different widths. First,
the thick line presents the result without state dependence calculated
as in Ref.~\cite{brack} associated with $pp\rightarrow d\pi^+$ using
the width
\begin{equation}
\Gamma = \frac{2}{3}\, \frac{{f^\ast}^2}{4\pi}\, \frac{q^3}{\mu^2}
\end{equation}
with $\mu$ and $q$ the pion mass and momentum. Ref.~\cite{brack}
adopted $f^\ast = 2 f$ from Chew-Low theory
and the $\pi NN$ coupling $f^2/4\pi = 0.08$ giving $\Gamma = 114$~MeV
at this energy. For the $\Delta$ mass the real part of the position of
the pole 20~MeV below the nominal $\Delta$ mass is used in the present
calculations. The normal solid line, actually indistinguishable from 
the thick one, is the result using the $^5S_2(N\Delta)$ width 78~MeV 
from Ref.~\cite{delta2}. Because in the proximity of the $\Delta$ 
threshold the widths are much larger than $|E-\Delta|$, the thick and 
normal solid lines are both very close to 90 degrees.
Further from the $\Delta-N$ mass difference the lines would be distinguishable. The dashed line
has a much smaller width 11.5~MeV for $^5D_2(N\Delta)$ and also the 
negative sign of $V_2(r)$ is included.

\begin{table}[tb]
\caption{Expectation values of centrifugal (left columns) and 
internal radial kinetic energies (right columns) for three intermediate 
$NN$ laboratory energies in $N\Delta$ states (MeV). These should be
added to the $N\Delta$ mass to get the corresponding effective 
threshold masses.}
\begin{tabular}{|l|c|c|c|c|}
\hline
 $NN$ & $N\Delta$  & 400 MeV & 578 MeV & 800 MeV \\ \hline
\ $^1D_2$ \ &\ $^5S_2$ & 0 \quad \; 44 & 0 \quad\; 45 & 0 \quad\; 88 \\
   &\ $^5D_2$ & 133 \quad 192 & 147 \quad 209 & 112 \quad 186 \\
   &\ $^5G_2$ & 398 \quad 457 & 484 \quad 578 & 499 \quad 616 \\
 \hline  
\ $^3F_3$ & $^5P_3$ & 33 \quad 67  & 36 \quad 68 & 39 \quad 113 \\
   & $^5F_3$ & 157 \quad 198 & 189 \quad 237 & 165 \quad 222  \\
 \hline  
\ $^1G_4$ & $^5D_4$ & 59 \quad 89 & 65 \quad 94  & 64 \quad 131  \\
   & $^5G_4$ & 173 \quad 207 & 214 \quad 257  & 214 \quad 264  \\
 \hline  
\ $^1S_0$ & $^5D_1$ & 298 \quad 438 & 331 \quad 509 & 365 \quad 580 \\ 
\hline
\ $^3P_1$ & $^5P_1$ & 53 \quad 111  & 52 \quad 100 & 47 \quad 133 \\
   & $^3P_1$ & 100 \quad 178  & 77 \quad 100 & 59 \quad 147 \\
   & $^5F_1$ & 262 \quad 340  & 308 \quad 408 & 288 \quad 382 \\ 
 \hline  
\ $^3F_2$ & $^5P_2$ & 38 \quad 72  & 36 \quad 66 & 33 \quad 110 \\
   & $^3P_2$ & 50 \quad 76   & 43 \quad 70 & 41 \quad 113 \\
   & $^5F_2$ & 172 \quad 212  & 188 \quad 234 & 151 \quad 212 \\ 
   & $^3F_2$ & 124 \quad 177  & 151 \quad 215 & 146 \quad 214 \\ 
 \hline  
\ $^3P_2$ & $^5P_2$ & 102 \quad 184  & 104 \quad 181 & 92 \quad 206 \\
   & $^3P_2$ & 430 \quad 561  & 293 \quad 388 & 147 \quad 249 \\
   & $^5F_2$ & 484 \quad 532  & 595 \quad 686 & 548 \quad 686 \\ 
   & $^3F_2$ & 425 \quad 509  & 518 \quad 657 & 589 \quad 791 \\   
 \hline
\end{tabular}
\label{ecents}
\end{table}

\subsection{Effective thresholds}
Finally, the expectation values of the centrifugal barrier
$\hbar^2 / (2m'r^2)$ are calculated for the three energies discussed and 
several $N\Delta$ states. From the results shown in Figs.~\ref{absdiba} 
and~\ref{nondibar}, apart from the factor $L'(L'+1)$, one would 
anticipate these to be rather similar within each $NN$ state. 
One can deduce the average centrifugal energies in each state
explicitly as presented in Table~\ref{ecents}, to be added to the
mass difference $\Delta - M$ to get an effective threshold. And really,
for the $N\Delta$'s coupled to $^1D_2(NN)$ at 578~MeV the expectation 
value $\langle 1/r^2 \rangle$ is 0.67~fm$^{-2}$ for $^5D_2(N\Delta)$, 
while for $^5G_2(N\Delta)$ it is 0.68~fm$^{-2}$, the same within 2\%.
(The reduced mass to give $\hbar^2 /(2m')=36.54\, {\rm MeV fm^2}$ has
been used.) 

However, there is a slight but larger dependence on energy as can be
seen following the horizontal rows, but this is not particularly
systematic. The most systematic dependence of the centrifugal energy
(apart from the factor $L'(L'+1)$) is on the angular momentum $L$
of the initial $NN$ state calculated for a fixed $N\Delta$ angular 
momentum $L'$: the larger $L$, the smaller $E_{\rm cent}$. 
This resembles the trend quoted earlier and seen in 
Ref.~\cite{dibarmass}: decrease of the orbital angular momentum 
in the transition $NN\rightarrow N\Delta$ is favoured. The 
effective threshold is then clearly lower than for $L' \ge L$. 
One can perhaps recognize the likeness
of the behavior with a soft rotator: the larger angular momentum
stretches the rotator increasing its moment of inertia and
decreasing the related energy quanta. However, here the 
``stretching'' angular momentum is mainly associated with the 
external $NN$ state, not the internal $N\Delta$. For higher angular 
momenta the incident nucleons remain further and the transition is 
more peripheral leading also to a larger average distance between 
the nucleon and the $\Delta$. In contrast, within the $N\Delta$
configuration the short-range $r^{-2}$ repulsion is also counteracted 
by a higher barrier at larger distances deepening the classically 
forbidden region and damping the tunneling into asymptotic ranges.
The $r^{-2}$ dependence has longer range than the strong interaction
or the extent of the confined wave function.
The situation may be compared with e.g. hydrogen atom states, where 
the electron probability density is not strongly pushed to asymptotic
regions by the centrifugal barrier - only rather the short-range
behavior is affected. 
So e.g. in the $^5D_4(N\Delta)$ state 
(coupled to $^1G_4(NN)$) the above expectation value 
$\langle 1/r^2 \rangle$ is 0.27~fm$^{-2}$, while in the 
$^5G_4(N\Delta)$ it is nearly the same 0.24~fm$^{-2}$ but nevertheless 
$\approx$10\% smaller. However, in comparison with the 
$^1D_2(NN)$ initial state the difference is towards even qualitatively smaller centrifugal energies, i.e. towards
larger $r^2$, with a factor of $\approx$2 between the two.

The expectation value of the kinetic energy associated with the 
radial degree of freedom also
increases with angular momentum $L'$ but most strongly with energy
above the nominal $N\Delta$ threshold. 
However, below, say $E({\rm lab}) \approx 600$ MeV, this is relatively
constant and, added together with the centrifugal energy
to the $N\Delta$ mass difference, might be 
considered to imply some kind of an effective threshold.  
Its nearly linear dependence above the nominal threshold 
apparently means that much of the excess $NN$ energy 
can be seen to emerge in this way within the $N\Delta$ system. 
The increasing kinetic energy means more curvature of the wave function
and the nodes (and maxima) coming closer to zero as was seen in 
Figs. ~\ref{absdiba} and~\ref{nondibar} after 600 MeV.

\begin{figure}[tb]
\includegraphics[width=\columnwidth]{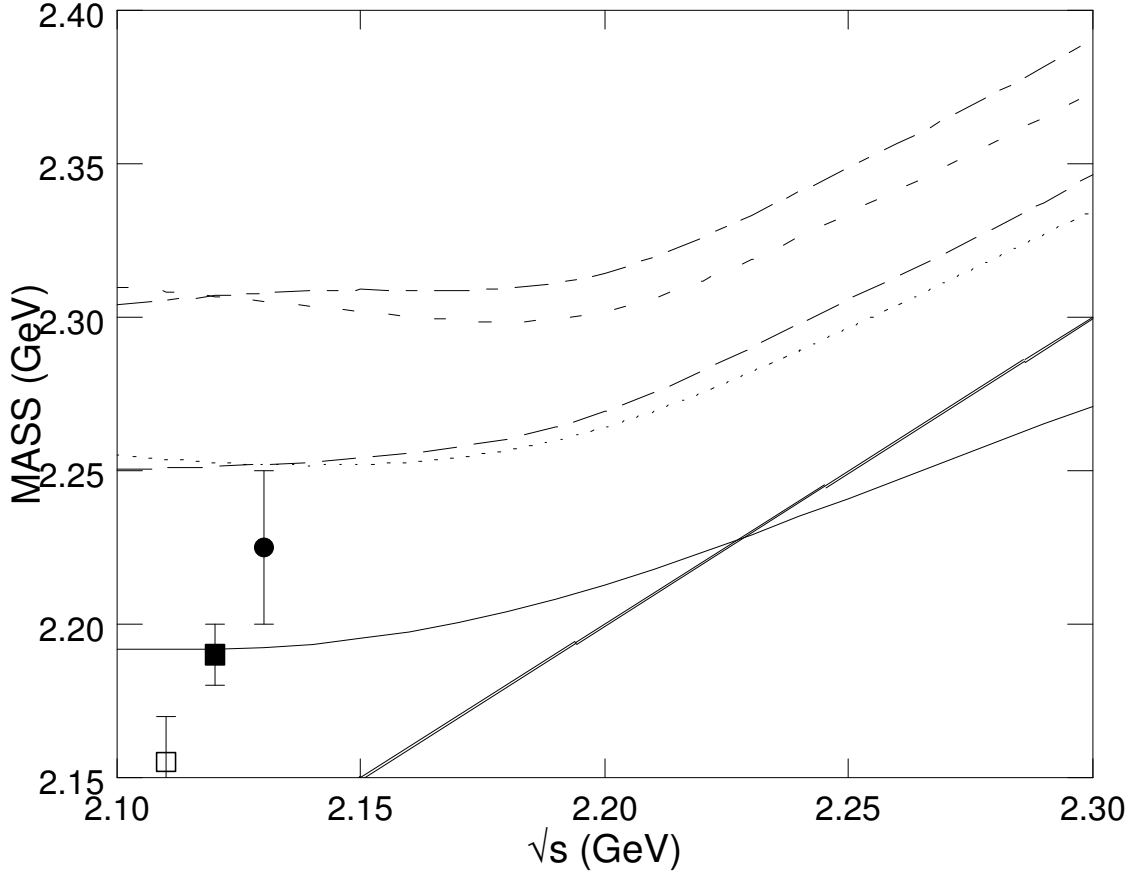}
\caption{The $N\Delta$ mass combined with the expectation value of the
kinetic energy components (centrifugal and radial)
for $^5S_2$ (solid), $^5P_3$ (long dashes),
$^5D_4$ (dash-dot), $^5P_2$ from $^3F_2(NN)$ (dotted) and $^5P_1$ 
(short dashes). The thick line shows the total energy $\sqrt{s}$ also
on the mass scale. The points indicate three dibaryon masses suggested 
in Ref.~\cite{yokosawa}.} \label{thresholds}
\end{figure}

Fig. \ref{thresholds} represents the effective thresholds thus 
calculated for the lowest energy $N\Delta$ admixture components, shown 
in Table~\ref{ecents},
as functions of the total center-of-mass energy $\sqrt s$. 
The lowest, solid line would be the lightest one, $^5S_2(N\Delta)$,
associated with the $^1D_2(NN)$ initial state. 
Compared with the lowest energy dibaryon suggested by 
Yokosawa~\cite{yokosawa} (hollow square) this threshold would need
some 20--40 MeV attraction. However, this requirement would conform 
very well with the early estimates for the $N\Delta$ binding energy 
of Arenh\"ovel \cite{arenbinding} (actually calculated for isospin 2
to avoid coupling to the always open $NN$ decay channels with isospin 1). 
A downward shift of the $N\Delta$ peak by about 20 MeV is 
also found in photodisintegration of deuterons in Ref.~\cite{wilh}, if
pion exchange interaction is added between the nucleon and the $\Delta$.
Therefore, $N\Delta$ attraction might allow the solid line to 
be accommodated with the mass range 2.14--2.17 MeV.
The long dashes present the threshold of the $^5P_3(N\Delta)$ 
component arising from
$^3F_3(NN)$. This is quite well established and the shoulder in its $NN$
phase shift is very well described by the isobar coupling \cite{delta2}.
Also its important role in the successful
 description of polarization phenomena
as well as differential cross section in $pp\rightarrow d\pi^+$ was first 
stressed in Ref.~\cite{ppdpi}. Now, its energy conforms rather well 
with the suggestion of Yokosawa as the possible $^3F_3$ 
dibaryon resonance at 2.20--2.25 MeV~\cite{yokosawa} (filled circle). 
Both of these points also agree well with Hoshizaki's $^1D_2(2.17)$ 
and $^3F_3(2.22)$ diproton resonances~\cite{hoshi60,hoshi61}.
The filled square indicates also Yokosawa's suggestion for a 
further possible triplet $P$ dibaryon (with a question mark) at
2.18--2.20 MeV. For this kind of
low threshold the present calculation would indicate rather as the 
starting $NN$ state $^3F_2$ (dotted curve) but still significantly 
higher than Ref.~\cite{yokosawa} (by $\approx$60 MeV). The $^3P_1$ 
initial state (short dashes), in turn, would yield still about 50~MeV 
more overestimate and $^3P_2$ still much more as seen from 
Table~\ref{ecents}.

Also indicated by the thick line in Fig.~\ref{thresholds} is the
c.m. energy itself on the mass axis. Namely for a resonance this 
line should cross the threshold or resonance mass curve. This does
happen for the $^5S_2(N\Delta)$ case, actually at the same energy as 
the calculated $^1D_2(NN)$ Argand diagram line crosses the y-axis in 
Fig.~5 of Ref.~\cite{delta2}. Therefore, in this respect the
$^5S_2(N\Delta)$ threshold effect resembles a
true resonance. However, in other cases the excess energy above threshold
in the $N\Delta$ states causes the crossing point to escape. Consequently,
the Argand diagram of the  $^3F_3(NN)$ ``dibaryon'' stays on the left 
side of the y-axis in that figure, although the corresponding shoulder
of the phase shift
in Fig.~6 (Ref.~\cite{delta2}) mimics well a resonance. Both diagrams 
have exactly the same behavior as found in energy-dependent and 
energy-independent phase shift analyses by Arndt 
{\it et al.}~\cite{arndt}.

There are two higher energy dibaryon suggestions by 
Yokosawa, both at 2.43--2.50~GeV ($^1G_4$ and a triplet state).
The dash-dot curve shows the described calculation for the former as too
low by over a hundred MeV as well as a candidate $^3P_1(NN)$
 also for the latter one
(short dashes). However, due to the energy dependence of the 
calculated expectation values the low-end threshold results may 
not be totally relevant in the case of higher energies where the
effective threshold is also larger and rapidly increasing.

\begin{figure}[tb]
\includegraphics[width=\columnwidth]{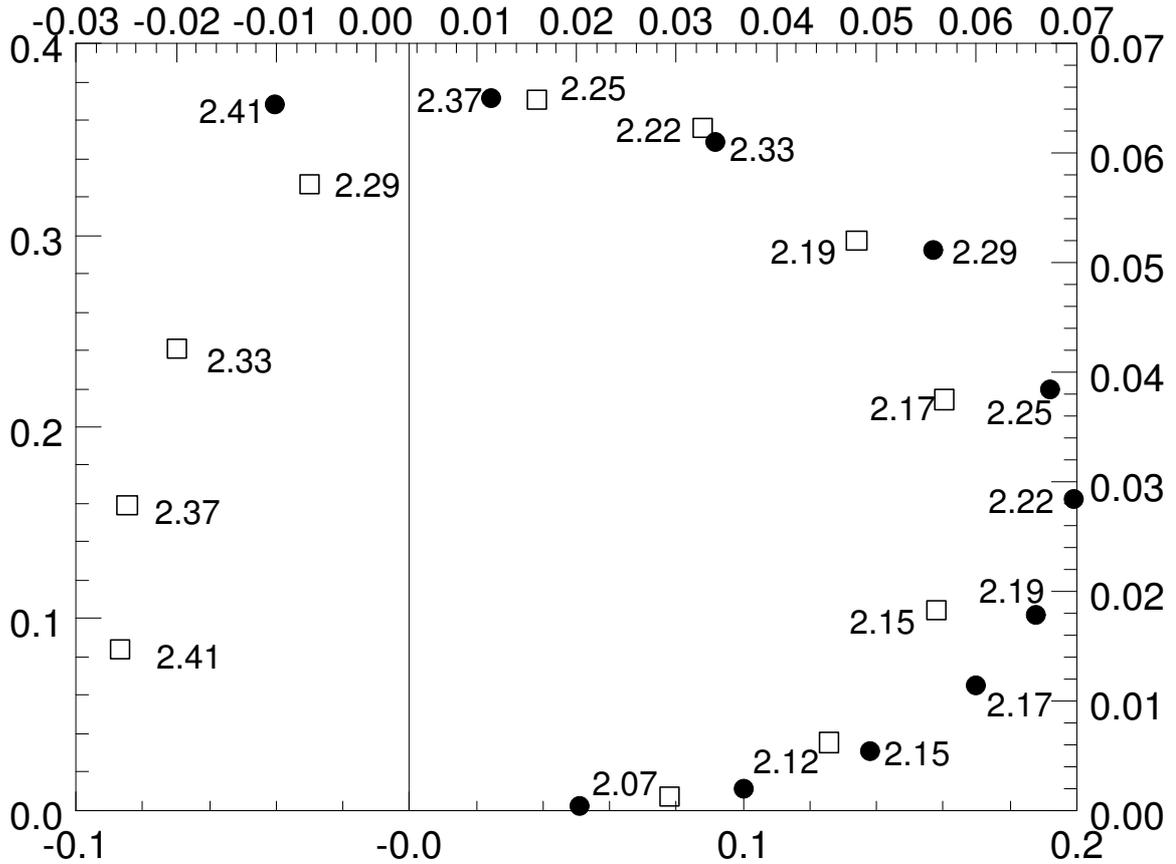}
\caption{Circles: $^5D_4(N\Delta)$ wave function at 1.5 fm 
for different c.m. energies (in GeV) imitating an Argand plot
(left and bottom scale). The overlap integral discussed in the text
(in ${\rm fm}^{1/2}$) shown by squares in top and right scale.
The imaginary axis (real part zero point) is given for the former.
Unessential overall 
minus sign is omitted.} \label{g4argand}
\end{figure}

Though it seems from Fig. \ref{thresholds} that in the 
$^5D_4(N\Delta)$ state the threshold cannot be reached 
and crossed, it may still be of some 
interest to study the behavior of the wave functions more directly
for qualitative insights. In the context of Figs. \ref{absdiba} and
\ref{nondibar} it was seen that the maxima of the wave functions for
$L' > 0$ could still continue growing beyond 800 MeV laboratory energy
(beyond 2.25 GeV mass). So even if this crossing does not take place,
there might appear some wide resonance-like peaking. 
Fig.~\ref{g4argand} shows this $N\Delta$ wave function at $r=$1.5 fm
as an Argand diagram arrangement (circles). 
At this distance the absolute
value of the wave function at 1100 MeV energy has its maximum. This
is also the energy at which the peak (outside Fig. \ref{absdiba}) levels
to a very wide global maximum $\approx$0.37. The idea is that the 
weighted wave function at some optimized distance
would be qualitatively representative for the behavior of 
the corresponding amplitude. The archetypal circle is featured, which
crosses the imaginary axis (drawn for this quantity) approximately 
for 2.39 GeV. So, even though
the c.m. energy does not formally reach the escaping
effective threshold (2.45 GeV at this incident energy) from 
Fig.~\ref{thresholds}, the corresponding 
amplitude can still have some resonance-like behavior reasonably
close to Yokosawa's experimental suggestion 2.43 MeV.
Further, shown by the squares is the overlap integral
\begin{equation}
I(^5D_4(N\Delta)) = \int_0^\infty w_D(r) j_0(qr/2) w_{N\Delta}(r)\, dr
\end{equation}
appearing in $f$-wave pion production in reaction
$pp\rightarrow d\pi^+$ (here $w_D(r)$ is the $D$-wave component 
of the deuteron and $q$ the center-of-mass momentum of the pion).
A similar resonance circle emerges. However,
the spherical Bessel function stresses the shorter ranges and
so this overlap crosses the imaginary axis at a lower energy than
the representative fixed-point value of the wave function.

\section{Conclusion}
In this work the effect of the centrifugal barrier in the $N\Delta$
configurations coupled to initial $NN$ states has been considered 
from different angles. First the effect on the magnitude of the wave
function as a function of $r$ is
studied for different angular momentum situations. It
was seen that, contrary to the first expectation, the angular
momentum $L'$ of the $N\Delta$ system itself has very little effect
on the shape of the wave function, only on the overall magnitude
with higher repulsions decreasing the probability of high $L'$
$N\Delta$ admixtures. This agrees with the finding of a more
phenomenological calculation \cite{dibarmass}. However, the initial
$NN$ angular momentum $L$ has a more significant influence in pushing 
the $N\Delta$ system apart.

Next the association of the state dependent width as a uniform
imaginary potential with the phase of the $N\Delta$ admixture is 
studied comparing the coupled channels results with simpler models
of separable wave functions. Significant differences are seen 
especially considering that normally such models do not include 
angular momentum dependence of the width implied in self-consistent 
coupled channels calculations \cite{delta2}.

It is also suggested that, due to the fact that in the 
complex potential now the $N\Delta$ wave function is 
confined producing finite expectation values for the 
centrifugal barrier and kinetic energy, it
is possible to define effective thresholds higher than the nominal
mass barrier $M_\Delta - M_N$ for different $N\Delta$ components.
This explains some parts of the wave function behavior seen above. 
Due to kinetic energy being assimilated in the configurations, 
these thresholds have
strong energy dependence above the nominal mass difference. In spite
of the thresholds escaping higher and higher with increasing energy
it was possible to see some resemblance to resonant behavior in the
$N\Delta$ wave functions and the transition amplitudes as exemplified 
in Fig. \ref{g4argand} for the $^5D_4(N\Delta)$ component originating
from $^1G_4(NN)$. It may be noted that these results can be regarded
as kinematic consequences. Any strong interaction model between the
nucleon and the $\Delta$ is not attempted here.

Apart from slight deviations these findings are qualitatively
in line with those of Ref.~\cite{dibarmass}. The difference of
principle is that in Ref.~\cite{dibarmass} the phase equivalence 
was forced to the interaction with and without the $N\Delta$ coupling,
while here no such constraint is explicitly imposed though the phase
shifts of e.g. Arndt {\it et al.}~\cite{arndt} are well reproduced. 
Of course, that constraint implicitly incorporates also the finite 
kinetic energy of the $N\Delta$ system.
To counteract the strongly attractive $N\Delta$ box and its 
iterations by repulsive $[V_2(r)]^2/\Delta E$ the denominator $\Delta E$
needs to become even smaller than $\Delta - N$ mass difference
(in the absence of the centrifugal barrier in $^5S_2(N\Delta)$
state). By definition, this is not possible in the present work.
Similar systematics holds also for the triplet states. In general, 
this energy denominator remains thus somewhat smaller than the 
presently calculated effective thresholds. Numerically, apparently the
implicit combination of the centrifugal energy and the $L'$ dependent
radial kinetic energy gave larger quanta of 40 MeV attributed to the
centrifugal part in Ref. \cite{dibarmass} as the overall 
rotational series
${\rm const} + 40 L'(L'+1)$ MeV. A quick look at Table~\ref{ecents}
confirms this simplified prescription as well valid for the $^1D_2$ 
and $^3P_1$ $NN$ states, but only qualitatively elsewhere with 
smaller energy quanta $\approx$30 MeV.

Numerically the effective thresholds for the lowest $L'$ $N\Delta$
states shown in Table \ref{ecents} and Fig. \ref{thresholds} agree 
relatively well with the suggested isospin one $^1D_2$ and $^3F_3$
dibaryons. Also the state $^1G_4$ can get some qualitative illumination
in terms of $N\Delta$ wave functions. Moreover the widths are agreeable
at relevant masses as shown earlier e.g. in Ref. \cite{delta2}.

\end{document}